\documentclass{article}

\usepackage{graphicx}
\usepackage{psfig}
\usepackage{epsfig}
\usepackage[round]{natbib}

\title{
Statistical modelling of tropical cyclone genesis:
a non-parametric model for the annual distribution
}

\begin{document}

\author{Tim Hall, GISS\footnote{\emph{Correspondence address}: Email: \texttt{tmh1@columbia.edu}}\\and\\
Stephen Jewson\\}

\maketitle

\begin{abstract}
As part of a project to develop more accurate estimates of the risks due
to tropical cyclones, we describe a non-parametric method for the statistical simulation of
the location of tropical cyclone genesis.
The method avoids the use of arbitrary grid boxes, and the spatial smoothing of the historical
data is constructed optimally according to a clearly defined merit function.
\end{abstract}

\section{Introduction}

We are interested in developing more accurate methods for the estimation of the various
risks associated with tropical cyclones, such as the risks of extreme winds, extreme rainfall
and extreme storm surge.
The methods for tropical cyclone risk assessment described in the scientific and engineering literature
can be categorised into \emph{local} methods and \emph{basin-wide} methods.
Local methods estimate the risk of high winds, rain or surge at a coastal location
using information from historical tropical cyclone events that made landfall near that location.
Basin-wide methods estimate risks using a model for the entire life-cycle of tropical
cyclones, from genesis to lysis. There are also methods in between, that model a part of the life-cycle
of tropical cyclones.
Some examples of local methods are described in~\citet{jagger} and~\citet{murnane},
and some examples of basin-wide methods are described in~\citet{drayton00}, \citet{fujii},
\citet{vickery00}, \citet{emanuel05}, \citet{darling91} and \citet{chu98}.
The basin-wide methods, \emph{if they can be made to work well},
are potentially the most accurate and the most useful.
They are potentially the most accurate because they make use of
all the available historical data,
and they are potentially the most useful because they can give estimates of
any of the various risks associated with tropical cyclones.
For instance a basin-wide model could be used to estimate the risk of a tropical
cyclone occurring at any point in the basin, not only over land: this
could be useful for shipping and the offshore oil industry.
And it could also be used to estimate the risk
of a tropical cyclone having an impact on more than two locations during its lifetime, such as
the risk of a hurricane hitting both Puerto Rico and Florida: this could be useful
for insurance companies that may sell insurance in both places.
Basin-wide models also have another advantage, which is that they can accommodate the inclusion
of weather, seasonal and year-ahead forecasts more easily than local models.

Given these potential advantages of basin-wide tropical cyclone risk models, and because of various
shortcomings in the basin-wide methods described in the literature,
we have initiated a project to build a new basin-wide tropical cyclone
model from scratch. One of the features of the model is that we
are paying great attention to the use of carefully designed statistical
procedures and methodologies. In particular:

\begin{itemize}

    \item By using non-parametric statistical methods we avoid the use of arbitrary grid boxes within the basin.
    The modelled properties of tropical cyclones are allowed to vary smoothly in space (and time),
    as they presumably do in reality.

    \item We use a merit function (the likelihood) that allows us to perform an objective
    comparison among different models

    \item All lengthscales and timescales used to select the data used in the model
    are derived optimally according to the merit function

    \item We evaluate our merit function in out-of-sample tests, to avoid overfitting and to account
    correctly for parameter uncertainty.

    \item We start with simple methods, and build up to more complex methods, again to avoid overfitting.

\end{itemize}

Following the various methods described in the literature (and cited above),
we divide the problem of basin-wide tropical cyclone modelling into various steps:
\begin{enumerate}
    \item Modelling annual rates
    \item Modelling the distribution of genesis in space and time
    \item Modelling tracks
    \item Modelling intensity along tracks, including lysis
    \item Modelling wind fields
    \item Modelling storm surge
    \item Modelling rainfall
\end{enumerate}

So far we have only considered step 3, the modelling of the shape of tropical cyclone tracks
(see~\citet{hallj05a}, \citet{hallj05b} and \citet{hallj05c}).
We have now turned our attention to step 2, the modelling of genesis, and that is the subject of this article.
A first look at step 1 is described in~\citet{j81}.

\section{Data}

As in our track modelling we are focussing initially on the Atlantic basin,
and the data we use is the `offical' National Hurricane Centre
track data set, known as HURDAT~\citep{hurdat}. We only use data from 1950, since this is the only
data that we consider to be sufficiently reliable.
Reliability increases from 1950 onwards because
starting from 1950 doppler radar was routinely used to determine wind speed.
HURDAT data from 1950 to 2004 contains 524 tropical cyclones, and
these are the data that we will use. Each storm has a unique data genesis point, and these 524
genesis locations are the input for our statistical model for genesis.

\section{Method}

As discussed in the introduction, our aim is to build models that possess a number of desirable
features.
The genesis model that we now describe, that possesses these features, is a two-dimensional kernel density with the
bandwidth fitted using cross-validation. The two dimensions are longitude and latitude:
at this point we ignore variations in genesis by season or by year, although we intend to consider
these in a later study.
The model gives a probability density $f(x,y)$ for tropical cyclone genesis at the point $(x,y)$ of:

\begin{equation}\label{density}
 f(x,y) = \frac{1}{N \sigma_x \sigma_y} \sum_{i=1}^{N} K \left(\frac{x-x_i}{\sigma_x},\frac{y-y_i}{\sigma_y} \right)
\end{equation}

\noindent where the $x_i$ and $y_i$ are the longitudes and latitudes of the historical genesis points,
$\sigma_x$ and $\sigma_y$ are bandwidths in the longitudinal and latitudinal directions, and
$K$ is a kernel function.
Large values for the bandwidths create a very smoothed density and small values create a very
multi-modal density.

\noindent The kernel $K$ must satisfy
\begin{equation}
 \int \int K(x,y) dx dy = 1,
\end{equation}

\noindent and the $\frac{1}{N}$ term ensures that
\begin{equation}
 \int \int f(x,y) dx dy=1,
\end{equation}
as it must for $f$ to be a probability density.

\noindent For convenience and simplicity we use a Gaussian kernel with $\sigma_x=\sigma_y$, and so:

\begin{equation}
 K(x,y)=\frac{1}{2 \pi} \mbox{exp} \left( -\frac{x^2+y^2}{2} \right)
\end{equation}

which gives:
\begin{equation}
 f(x,y) = \frac{1}{2 \pi N \sigma^2} \sum_{i=1}^{N}
          \mbox{exp} \left( -\frac{(x-x_i)^2+(y-y_i)^2}{2 \sigma^2}\right)
\end{equation}

The optimal bandwidth $\sigma$ is determined using a jack-knife cross-validation procedure as follows:

\begin{itemize}

    \item We loop over a range of values for the bandwidth

    \item We loop over the 55 years of data

    \item For each value of the bandwidth, for each year of data, and for each
    genesis point that occurs within that year, we calculate
    the density at that point using expression~\ref{density}, but eliminating
    all the data points in the same year from the sum

    \item For each value of the bandwidth, we calculate the likelihood score as
    the product of the densities at all genesis points

    \item We find the value of the bandwidth that gives the highest likelihood score

\end{itemize}

\section{Results}

\begin{figure}[!hb]
  \begin{center}
%    \scalebox{0.8}{\includegraphics{../timhall5/figs/pscore}}
    \scalebox{0.8}{\includegraphics{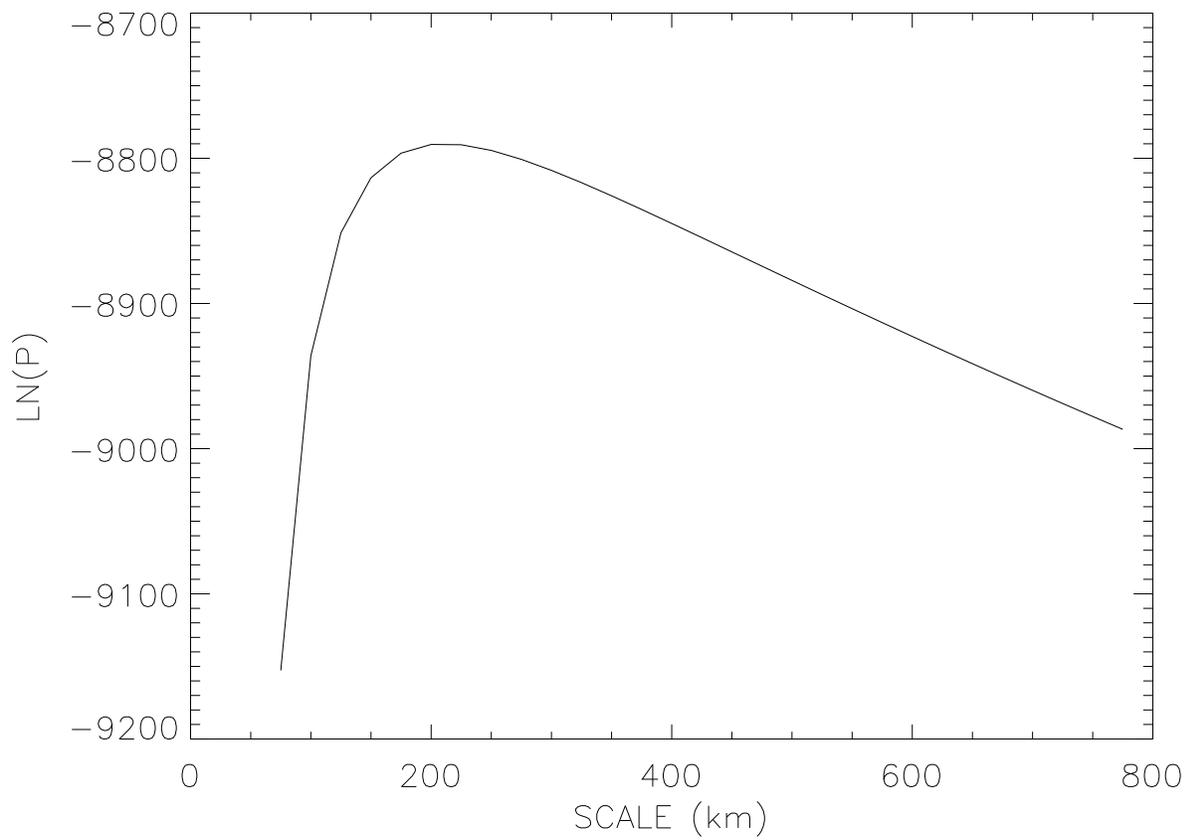}}
  \end{center}
    \caption{
The out-of-sample log-likelihood versus bandwidth for the kernel density model
described in the text. The maximum log-likelihood is at a bandwidth of 210km.
     }
     \label{f01}
\end{figure}

\begin{figure}[!hb]
  \begin{center}
%    \scalebox{0.8}{\includegraphics{../timhall5/figs/pdf_noseason_3scale}}
    \scalebox{0.8}{\includegraphics{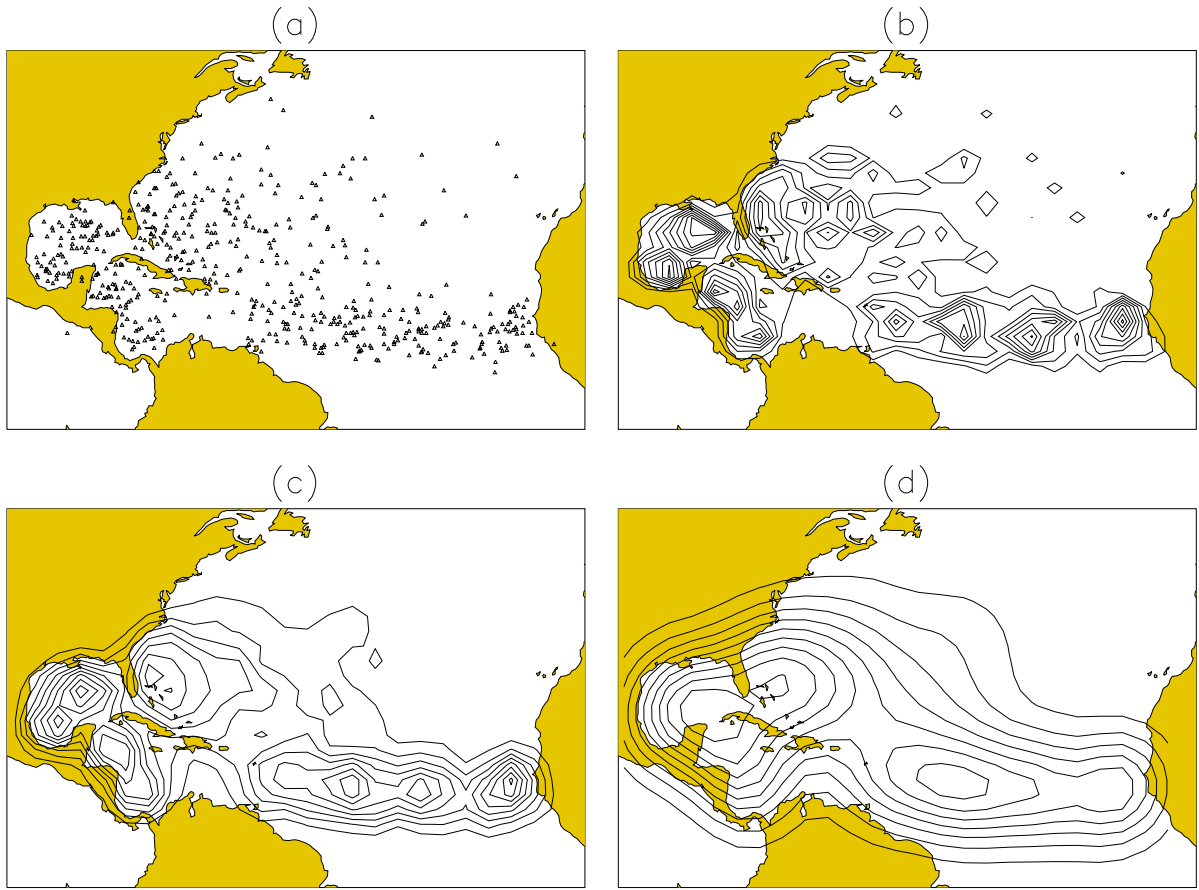}}
  \end{center}
    \caption{
Panel (a) shows the observed tropical cyclone genesis locations for the period 1950 to 2004 (524 points),
and panels (b), (c) and (d) show kernel densities estimated using bandwidths of 100km, 210km and 500km respectively.
     }
     \label{f02}
\end{figure}

We now show the results from the fitting of the kernel density to the observed genesis points.
The variation of the likelihood score with the bandwidth is shown in figure~\ref{f01}. There is
a very clear maximum of the likelihood function at a bandwidth of 210km.
Figure~\ref{f02} shows the historical hurricane genesis locations, and estimated densities based
on the kernel model.
Panel (a) shows the historical genesis locations,
panel (b) shows a density derived using a bandwidth of 100km,
panel (c) shows a density derived using the optimal bandwidth of 210km, and
panel (d) shows a density derived using a bandwidth of 500km.
The effects of undersmoothing and oversmoothing
can be seen very clearly in panels (b) and (d).

\subsection{Simulations}

Having fitted a probability density function to the observed hurricane genesis points, we can
now simulate as many hurricane genesis points as we desire. The simulation method we use works as follows:

\begin{itemize}
    \item We normalise the density $f(x,y)$ to have a maximum value of 1.
    \item We simulate random values of $(x,y)$ from a region that covers the entire domain. $x$ and $y$
    are simulated from independent uniform distributions.
    \item We then either accept or reject each simulated value of $(x,y)$ randomly, with a probability
    given by the normalised density.
\end{itemize}

\begin{figure}[!hb]
  \begin{center}
%    \scalebox{0.8}{\includegraphics{../timhall5/figs/gensim_noseason}}
    \scalebox{0.8}{\includegraphics{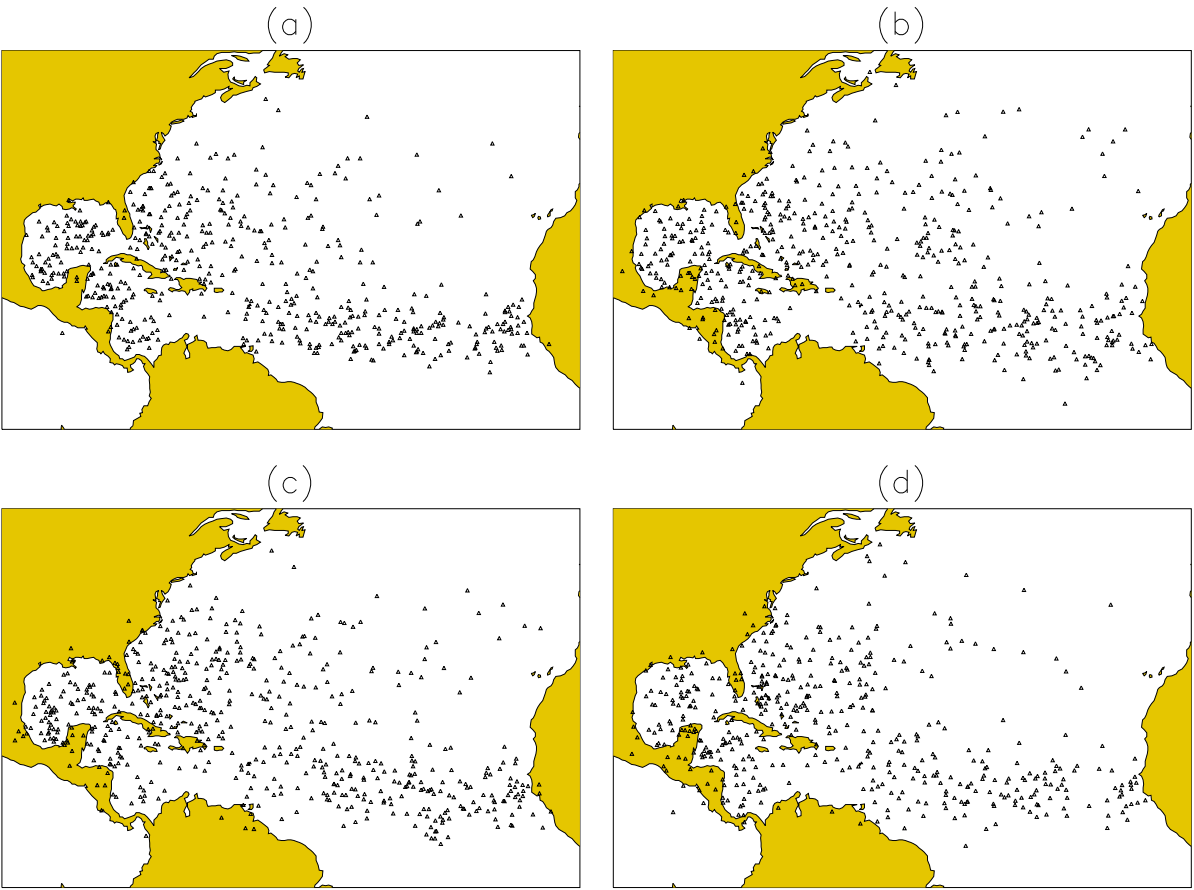}}
  \end{center}
    \caption{
Panel (a) shows the observed genesis locations for tropical cyclones in the period 1950 to 2004.
There are 524 points. Panels (b), (c) and (d) each show 524 simulated genesis locations from
the model described in the text.
     }
     \label{f03}
\end{figure}

Figure~\ref{f03} shows three realisations from such simulations,
each of 524 points (in panels (b), (c) and (d)) along with the 524 historical genesis points (in panel (a)).
We can see that the simulations follow the pattern of historical genesis reasonably closely, but
are different in detail, as we would expect.

One of the shortcomings of the model described above, apparent in figure~\ref{f03}, is that there are a number of genesis
points that have been simulated over land. This is non-physical in most cases, although there are occasional
genesis points over Florida and the Yucatan in the observations.
Non-physical genesis points in the simulations can be rejected to solve this problem.

\section{Conclusions}

We have described a statistical model for the location of hurricane genesis.
Our model is a non-parametric kernel density, with the bandwidth fitted using
a cross-validation procedure that optimises the out-of-sample likelihood.
The advantages of this approach include:
\begin{itemize}
    \item not having to define grid boxes
    \item the use of a clear merit function
    \item making the best use of the historical data by avoiding over- or under-fitting
\end{itemize}

This model is intended as the first model in a hierarchy,
and as such it can probably be beaten by a more complex model.
Its value is that it sets an initial standard.
Because of the use of a well-defined
merit function it will be easy to check whether the model has been beaten or not.

\begin{figure}[!hb]
  \begin{center}
%    \scalebox{0.8}{\includegraphics{../timhall5/figs/genobs}}
    \scalebox{0.8}{\includegraphics{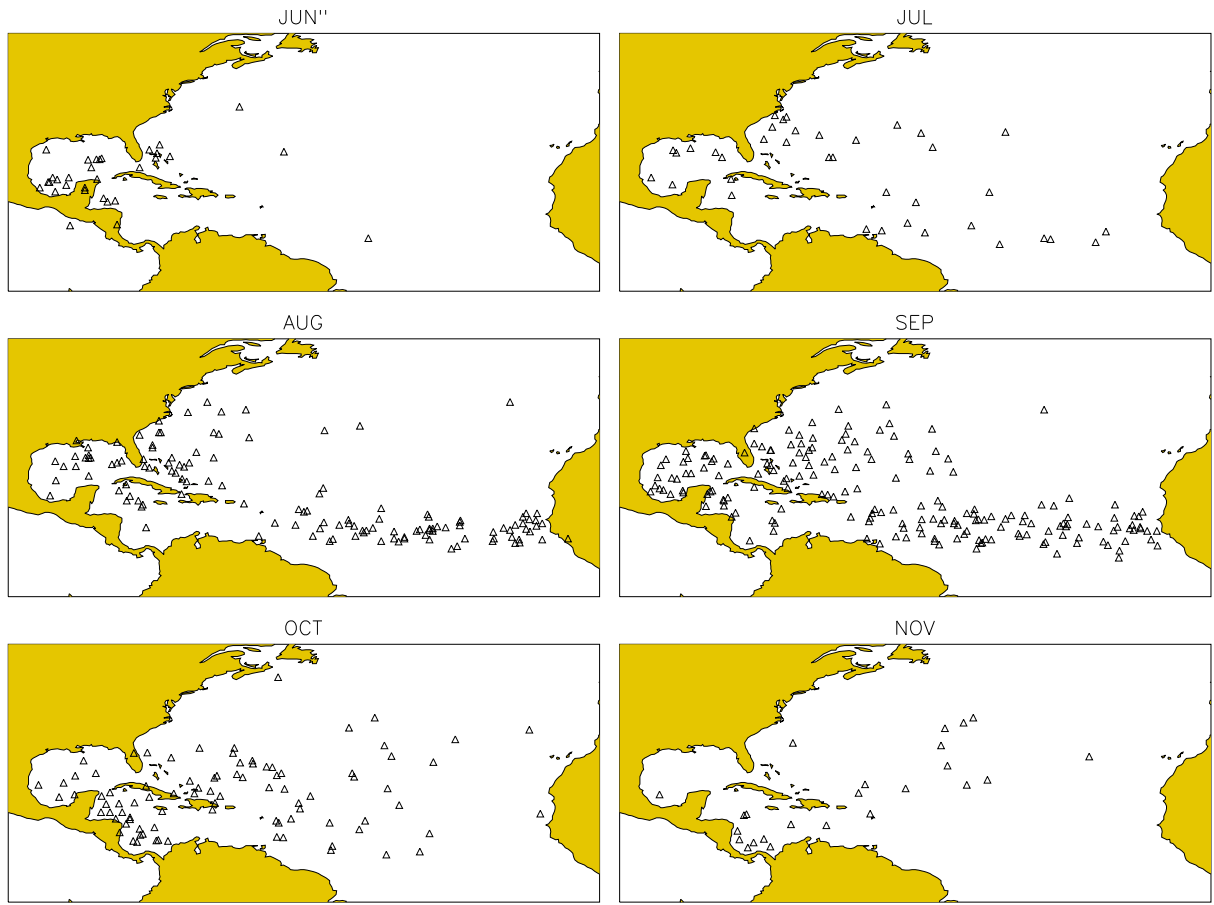}}
  \end{center}
    \caption{
The observed tropical cyclones genesis locations for the period 1950 to 2004, by month.
There is a clear variation of genesis location by month, suggesting that the model described
in the text could possibly be improved by including time as a third dimension in the kernel
density.
     }
     \label{f04}
\end{figure}

There are a number of ways that one might try to improve the performance of this model, such as:
\begin{itemize}

    \item The introduction of seasonality. Figure~\ref{f04} shows the observed genesis points by month.
    There are distinctively different patterns in each month.

    \item The introduction of different smoothing in the longitude and latitude directions.

    \item The use of kernels other than the gaussian kernel (although we have no particular reason to think
    that this will give an improvement, it may).

\end{itemize}

In~\citet{emanuel05} there is some discussion of issues related to whether the historical genesis data is accurate prior to
the introduction of satellite observations in the 1970s. If this is considered an important issue then
it would be easy to refit the current model but using only the more recent data.

\bibliography{timhall5}

%%%%%%%%%%%%%%%%%%%%%%%%%%%%%%%%%%%%%%%%%%%%%%%%%%%%%%%%%%%%%%%%%%%%%%%%%%%%%%%%%%%%%%%%%

\newpage

%%%%%%%%%%%%%%%%%%%%%%%%%%%%%%%%%%%%%%%%%%%%%%%%%%%%%%%%%%%%%%%%%%%%%%%%%%%%%%%%%%%%%%%%%

\newpage

%%%%%%%%%%%%%%%%%%%%%%%%%%%%%%%%%%%%%%%%%%%%%%%%%%%%%%%%%%%%%%%%%%%%%%%%%%%%%%%%%%%%%%%%%

\newpage

%%%%%%%%%%%%%%%%%%%%%%%%%%%%%%%%%%%%%%%%%%%%%%%%%%%%%%%%%%%%%%%%%%%%%%%%%%%%%%%%%%%%%%%%%

%%%%%%%%%%%%%%%%%%%%%%%%%%%%%%%%%%%%%%%%%%%%%%%%%%%%%%%%%%%%%%%%%%%%%%%%%%%%%%%%%%%%%%%%%

\newpage

\end{document}